\documentclass[preprint,3p,authoryear]{elsarticle}

\usepackage{graphics}
\usepackage{graphicx}
\usepackage{epsfig}

\usepackage{amssymb}

\journal{New Astronomy}

\begin{document}

\begin{frontmatter}



\title{A comprehensive study of six Algol type binaries}


\author[label1]{A. Liakos}
\ead{alliakos@phys.uoa.gr}
\author[label2]{P. Zasche}
\author[label1]{P. Niarchos}

\address[label1]{Department of Astrophysics, Astronomy and Mechanics, National and Kapodistrian University of Athens, GR 157 84 Zografos, Athens, Hellas}

\address[label2]{Astronomical Institute, Faculty of Mathematics and Physics, Charles University Prague, CZ-180 00 Praha 8, V Hole\v{s}ovi\v{c}k\'ach 2, Czech Republic}

\begin{abstract}
CCD light curves of the Algol type eclipsing binaries DP Cep, AL Gem, FG Gem, UU Leo, CF Tau and AW Vul were analysed using the Wilson-Deninney code and new geometric and absolute parameters were derived. Due to cyclic apparent orbital period changes of the systems, probably caused by the Light-Time Effect, the contribution of a third light was taken into account in the light curve solution. All the reliable timings of minima found in the literature were used to study the period variations and search for the presence of a tertiary component in the systems. A comparison between the parameters of the third body derived from the light curve and orbital period analyses is also discussed. Moreover, the absolute parameters of the eclipsing binary components were also used to determine their current evolutionary state.

\end{abstract}


\end{frontmatter}


\section{Introduction}
\label{1}

The main purpose of this work is the derivation of the light curve (hereafter LC) and orbital period parameters of each system, and a description of the components' current evolutionary stage. The systems were selected due to their \emph{\textbf{O}bserved -- \textbf{C}alculated} times of minima periodic variations (hereafter O -- C). The LC observations were performed in order: (a) to investigate the photometric presence of a tertiary component near the eclipsing binary (hereafter EB) via a third light contribution, (b) to make an approximate calculation of the absolute parameters of the components and (c) to search for possible pulsational behaviour in the systems which are candidates for including a $\delta$ Sct component.

According to the O -- C solution, it is feasible to detect which physical mechanisms play a role to the period modulation, such as a third body or the mass transfer between the components or mass loss from the system or magnetic activity. On the other hand, from the LC solution it is possible to determine the Roche geometry of the EB, such as semi-detached, detached or contact configuration. A plausible conclusion is that these solutions are qualitatively connected, even though the methods used for the derivation of their results are different. The LC solution is based on the model of the photometric variation of the EB, while the O -- C solution counts on the variability of its orbital period. Obviously, the LC is an instant image of the EB's long life, while the period changes correspond to a larger time scale. However, although the period variability duration of the system is small (order of decades) comparing with its total lifetime, one is able to unify the information provided from both analyses and obtain a consistent picture of the EB. The existence of a third body orbiting an eclipsing pair is not always easy to be proved. The tertiary component has to be of the same order of magnitude as the EB, in order to be detected photometrically through its light contribution to the total luminosity of the triple system. However, if the third body is an evolved star (e.g white dwarf) it is obvious that it will affect the EB's period, but not the LC. On the other hand, one can also imagine a different situation when only a weakly bound tertiary contributes a non-negligible amount of the third light, but no period modulation is detectable even on longer time scales.

Concluding, in the present study we selected the following six cases showing interesting variation of minima times in the O -- C diagram. The results derived from LC and O -- C analysis for each system and their probable connection will be discussed.

\subsection{Individual systems}
\textbf{DP Cep:} This system (V=12.9 mag, P$\sim$1.27 days) was discovered by \citet{b51} who determined its period. \citet{b52} obtained the first photographic LC of the EB, reported a more accurate period and, according to its B-V index, they listed its spectral type as F0. \citet{b08} classified the system as a semidetached with mass ratio of 0.6, and classified its components as F0 and G0 for primary and secondary, respectively. The first CCD LC was obtained by INTEGRAL (INTErnational Gamma-Ray Astrophysics Laboratory) mission \citep{b31} using the OMC (Optical Monitoring Camera), and \citet{b53} performed the first LC analysis.

\textbf{AL Gem:} The light variability of this system (V=9.77 mag, P$\sim$1.39 days) was reported by \citet{b01} and the first photoelectric LC was obtained by \citet{b02}. The system was classified as F4-F6 type star by \citet{b04}, while \citet{b03} measured its colour indexes. \citet{b05}, \citet{b06}, \citet{b22} and \citet{b21} calculated the absolute parameters of the system's components. The spectroscopic survey of \citet{b07} classified the EB as a F5 type, while \citet{b08} referred its primary and secondary components as F5 and K7 type stars, respectively. The first CCD LC was observed by the ASAS project \citep{b17}, but it is poorly covered. Finally, AL Gem is concerned as a candidate triple system \citep{b09}, but its hypothetical tertiary component has not been detected so far.

\textbf{FG Gem:} \citet{b10} discovered the variable nature of this EB (B=12.6 mag, P$\sim$0.82 days). Except for the observations for minima timings determination, no complete LC of the system existed until early the decade of 2000. The first LC was obtained by the ASAS project \citep{b17}, while \citet{b11} obtained B and R filter LCs and presented the first geometrical solution. The first approximate spectral classification, namely F7 for primary and K3 for secondary components, respectively, and the derivation of their absolute parameters were made by \citet{b21}. The spectral type of FG Gem is still uncertain since many catalogues refer different colour indexes estimates. In particular, \emph{Lick NPM2 Catalogue} \citep{b12}: B -- V=0.25, \emph{All-sky Compiled Catalogue of 2.5 million stars} \citep{b16}: B -- V=0.835 and \emph{Tycho-2 Catalogue} \citep{b14}: B -- V=0.02. In the catalogue of \emph{T$_{eff}$ and metallicities for Tycho-2 stars} \citep{b15}, except the B -- V indexes, the V -- K index and the metallicity were also taken into account for the classification. According to this catalogue,
although the errors are large, the most trustable temperature estimation of the system as T$\sim$7700 +4800/-590 K was provided, and, in addition, the distance of the system as 223 +283/-154 pc was estimated.

\textbf{UU Leo:} The eclipsing status of this system (B=12.1 mag, P$\sim$1.68 days) was mentioned by \citet{b18} and its period was given by \citet{b19}. Many timings of minima exist so far, but a complete LC was unavailable. \citet{b22} and \citet{b20} referred the system to be of A2 spectral type. The results of \citet{b21} showed that this result corresponds to the primary component, and in addition they found that the EB's secondary is of G1 spectral class. UU Leo is referred as candidate for triplicity \citep{b09} and also as candidate system for including a pulsating component \citep{b23}. \citet{b24} obtained an O -- C analysis of the EB, and concluded that the mass transfer between the components and magnetic phenomena are the responsible mechanisms for its orbital period changes.

\textbf{CF Tau:} Light variations of this EB (V=10.05 mag, P$\sim$2.76 days) were firstly announced by \citet{b25}, while \citet{b26} obtained the first visual LC. \citet{b27} published the first accurate ephemeris of the system, \citet{b03} derived its Str\"{o}mgen colour indexes and \citet{b22} calculated its absolute parameters based on the photometric parallaxes. \citet{b03} classified the system as G0 type, while the same spectral type is also referred by the \emph{Henry Draper Catalogue identifications for Tycho-2 stars} \citep{b29} and \emph{T$_{eff}$ and metallicities for Tycho-2 stars} \citep{b15} catalogue, where, in addition a distance of 159 +170/-79 pc was calculated. In agreement with this spectral values, \citet{b21} estimated the absolute parameters of the EB and classified its components as G0 and K2, for primary and secondary, respectively. \citet{b28} performed photoelectric UBV observations of the EB and calculated its colour indexes in various phases. LITE was proposed as the most plausible explanation for the period changes of the system according to the O -- C analysis of \citet{b30}. A partial
CCD LC of the EB by the ASAS \citep{b17} and a complete one by SWASP \citep{b32} projects were obtained.

\textbf{AW Vul:} The discovery of the system (V=11.1 mag, P$\sim$0.81 days) was made by \citet{b33}. \citet{b19} and \citet{b27} reported its first ephemerides. \citet{b22}, \citet{b20} and \citet{b21} calculated the absolute parameters of the system, and they are in agreement for the spectral type of the primary component as F0. The visual LC of the EB was obtained by \citet{b34}, while the only existed CCD LC is poorly covered by \citet{b35}. \citet{b36} classified the system also as F0 type and determined its distance as 440 pc. In addition, in the catalogue of \emph{T$_{eff}$ and metallicities for Tycho-2 stars} \citep{b15} the distance of the system is given as 155
+367/-277 pc. Finally, AW Vul is listed in the catalogue of possible triples \citep{b09} and its hotter component is candidate for pulsations \citep{b23}.

\section{Observations and data reduction}
\label{2}

The BVRI CCD photometric observations of AL Gem, UU Leo and AW Vul were carried out at the Gerostathopoulion Observatory of the University of Athens from January to July of 2010. The data of DP Cep (V-filter), FG Gem (V-filter) and CF Tau (unfiltered) were taken from the \emph{INTEGRAL/OMC} \citep{b31}, \emph{ASAS} \citep{b17} and \emph{SWASP} \citep{b32} projects, respectively.

Aperture photometry was applied to the data of AL Gem, UU Leo and AW Vul and differential magnitudes using the software \emph{MuniWin} v.1.1.26 \citep{b37} were obtained. The observational details are given in Table \ref{tab1}, where it is listed: The \emph{nights spent} and the \emph{time span} of the observations, the \emph{instrumentation} setup and the \emph{comparison} stars $C$ and $K$ used for each case.

\begin{table}[h]
\centering

\caption{The photometric observation log.}

\label{tab1}
\begin{tabular}{ccccc}
\hline
\textbf{System}   & \textbf{Nights} &  \textbf{Time span}   &      \textbf{Instrumentation}      & \textbf{Comparison}                    \\
                  & \textbf{spent}  & \textbf{(Day/Month)}  &      \textbf{(Telescope \& CCD)}   &    \textbf{stars}                      \\
\hline
AL Gem            &         14      &    31/01 -- 03/04     &     0.4m Cassegrain                &   C: TYC 1356-1240-1 (V=9.2 mag)       \\
                  &                 &                       &      \& ST--10 XME                 &   K: TYC 1356-0980-1 (V=10.8 mag)      \\
UU Leo            &         12      &    21/02 -- 08/03     &     0.4m Cassegrain                &   C: TYC 0834-0788-1 (B=11.6 mag)      \\
                  &                 &                       &      \& ST--10 XME                 &   K: TYC 0834-1173-1 (V=13.0 mag)      \\
AW Vul            &         8       &    02/07 -- 16/07     &   0.2m Newt. Reflector             &   C: TYC 2160-0945-1 (V=11.7 mag)      \\
                  &                 &                       &     \& ST--8 XMEI                  &   K: TYC 2160-0868-1 (V=11.1 mag)      \\

\hline

\end{tabular}
\end{table}

\section{Light curve analysis}
\label{3}

The complete LCs of each system were analysed using the \emph{PHOEBE} v.0.29d software \citep{b38} which uses the method of the 2003 version of the Wilson-Devinney (WD) code \citep{b39,b40,b41}. In the cases where data of various
photometric filters were available, the analysis was performed simultaneously in all LCs. Due to the absence of spectroscopic mass ratios, the `$q$-search' method using a step of 0.1 was trialled in Modes 2 (detached system), 4 (semi-detached system with the primary component filling its Roche Lobe) and 5 (semi-detached system with the secondary component filling its Roche Lobe) to find feasible (`photometric') estimates for the mass ratio $q_{ph}$. The latter, this value of mass ratio was set as initial input and treated as free parameter. The `Multiple Subsets' procedure was used to approach finally adopted solutions. The temperature values of the primaries were assigned values according to their spectral types by using the correlation given in the tables of \citet{b42} and were kept fixed during the analysis, while the temperatures of the secondaries $T_2$ were adjusted. The albedos, $A_1$ and $A_2$, and gravity darkening coefficients, $g_1$ and $g_2$, were set to the generally adopted values \citep{b13,b55,b56} for the given spectral types of the components. The linear limb darkening coefficients, $x_1$ and $x_2$, were taken from the tables of \citet{b43}; the dimensionless potentials $\Omega_{1}$ and $\Omega_{2}$, the fractional luminosity of the primary component $L_{1}$ and the inclination $i$ of the system's orbit were set in the programme as adjustable parameters. Due to a possible existence of a tertiary component orbiting each EB, suggested by the cyclic orbital period changes, the third light option $l_{3}$ was taken into account. In addition, in the cases where asymmetries on the LC were detected, the existence of a cool spot on the surface of the cooler component was assumed and its parameters, namely the latitude $Lat$, longitude $Long$, radius $R$ and temperature factor $T_f$ (=T$_{spot}$/T$_{surface}$) were set also as free parameters. For each case the best-fit model and observed LCs are presented in Figs 1-3 with corresponding parameters given in Table \ref{tab2}.

\subsection{DP Cep}

A value of 7300 K , typical for a F0 type star, was given to the primary component's temperature. Mode 2 was proved that provides the best fit for the data, and the mass ratio, through the q-search method, turned out as $\sim$0.7.

\subsection{AL Gem}

The effective temperature of the primary component $T_1$ was set to 6650 K according to its spectral type (F5V). In the LC analysis, the minimum sum of weighted squared residuals ($\sum res^{2}$), was found in Mode 2, where the $q$-search yielded a value of mass ratio around 0.2. A slight asymmetry before and after the secondary minimum led us to place a cool spot on the secondary component's surface. We preferred to characterize this component as magnetically active because the primary one, according to its spectral type, is very close to the limit of having radiative or convection zone, therefore the cooler one is better candidate for magnetic activity.

\subsection{FG Gem}

The most reliable spectral estimation was given by \citet{b15}, where the temperature of the system is referred as $\sim$7700 K, a value which was assigned to the primary's temperature. The minimum $\sum res^{2}$ was found in Mode 5, where the `q-search' method in this mode converged to a mass ratio value close to 0.4. Moreover, the data coverage and quality for this system is still rather poor, therefore the results may be not very convincing.

\subsection{UU Leo}

According to the literature (see section 1), an A2 spectral type for the primary component of the eclipsing pair was suggested, therefore $T_1$ was given a value of 9000 K. The best fit was achieved in Mode 2, where the mass ratio yielded a value of 0.3. Mode 5 seemed to describe the data also in a satisfactory way with almost the same value of mass ratio, however, the detached configuration, suggested as adopted solution, provided slightly smaller value of sum of squared residuals. Similarly to AL Gem, brightness irregularities on both sides of the secondary minimum suggested a presence of a cool spot on the solar-like component of the system, thus the final photometric solution includes its parameters as well.

\subsection{CF Tau}

A corresponding temperature ($T_1$=5950 K) for a G0 type star was assigned for the primary. The relatively small depth difference of the minima led the photometric search for the best $q_{ph}$ to turn out a value close to unity. Since the data were obtained without any standard photometric filter, the bolometric (linear) limb darkening coefficients were selected as the most realistic values.

\subsection{AW Vul}

A typical value of temperature for a F0 type star was set for the primary component, according to various classifications (see section 1). Mode 5 was found to describe the LC better, and the `q-search method' yielded a mass ratio of 0.5.


\begin{figure}
\centering
\begin{tabular}{cc}
\includegraphics[width=8cm]{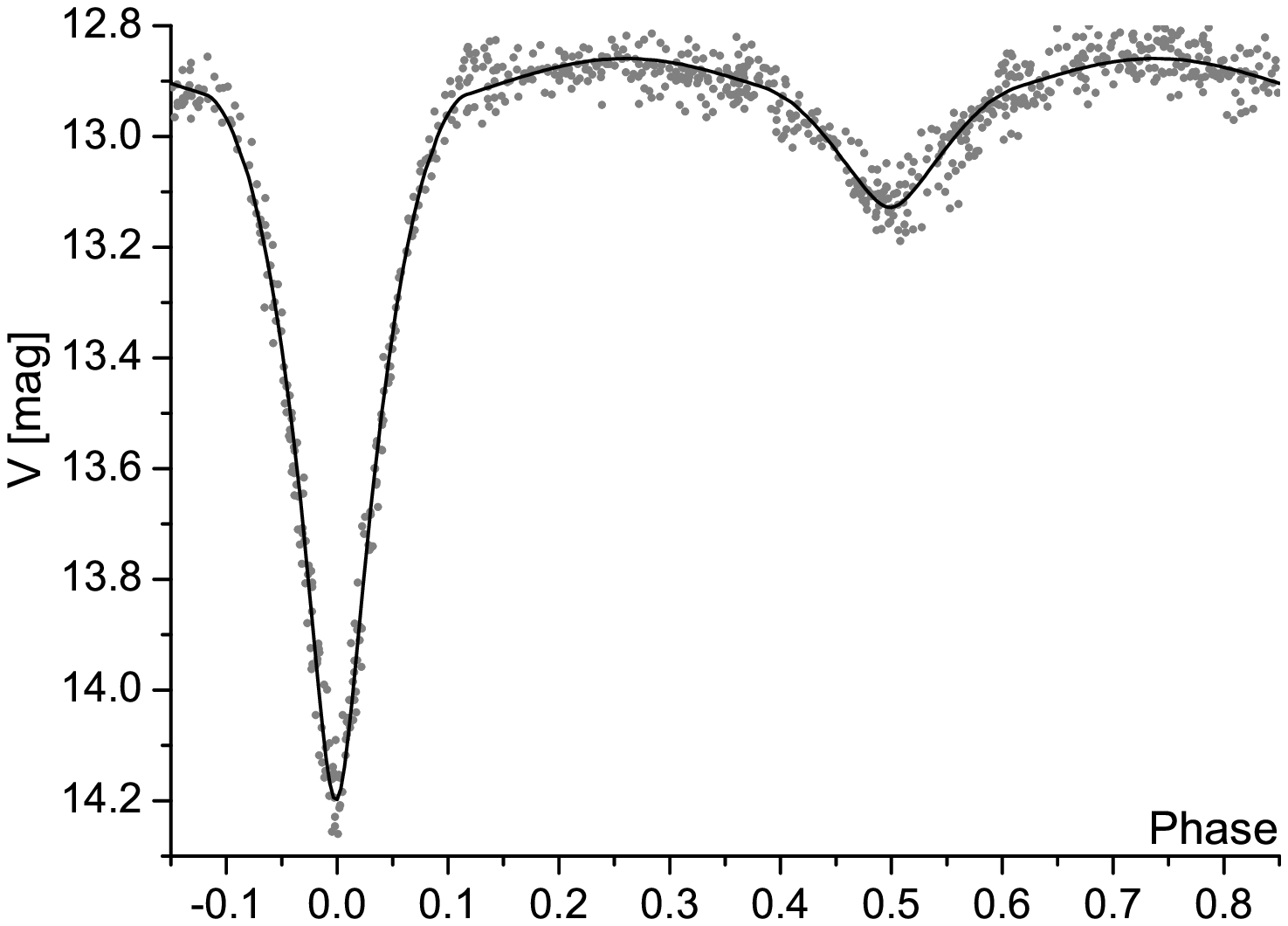}&\includegraphics[width=8cm]{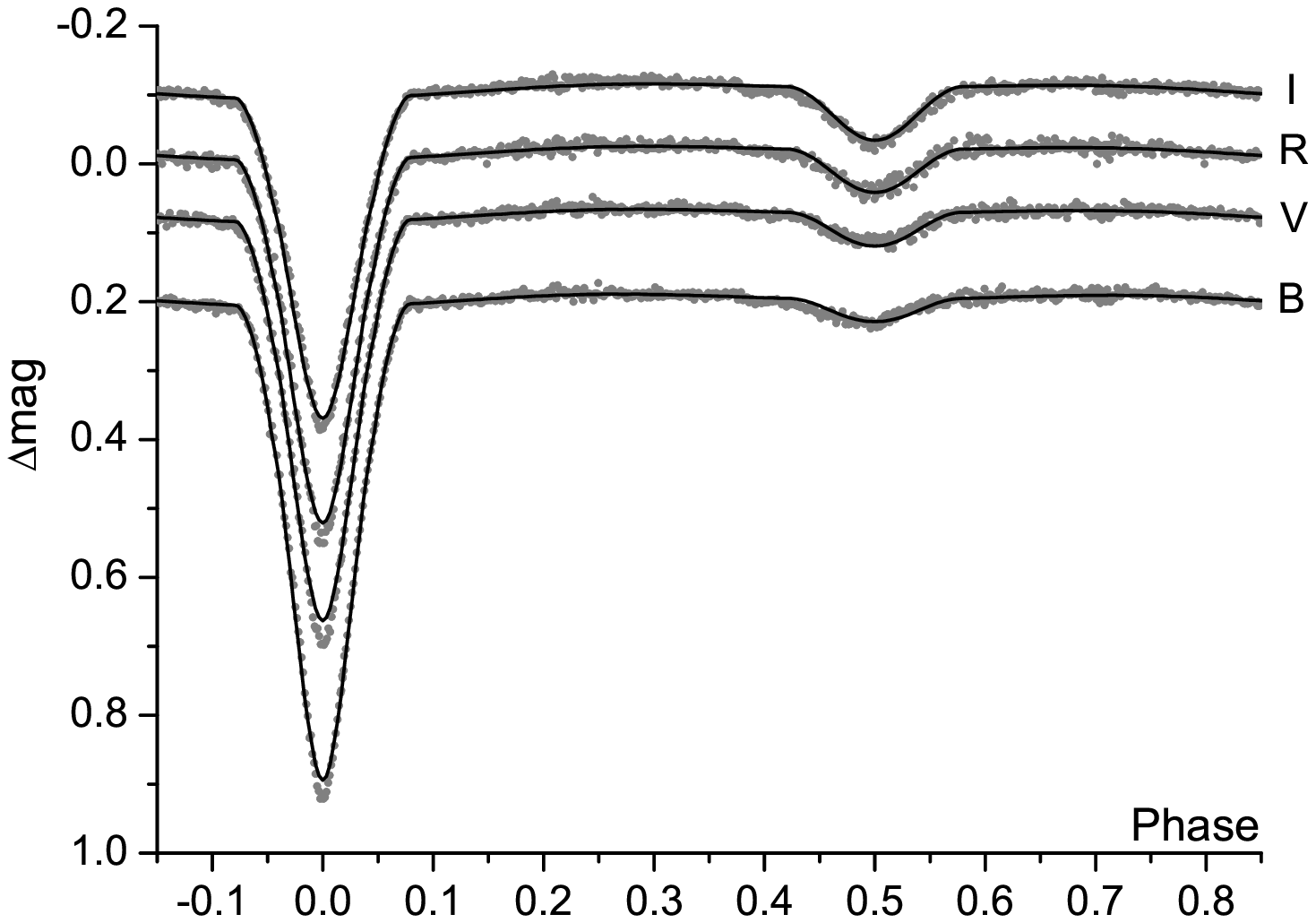}
\end{tabular}
\label{fig1} \caption{Theoretical (solid lines) and observed (grey points) LCs for DP Cep (left) and for AL Gem (right).}

\vspace{5mm}

\begin{tabular}{cc}
\includegraphics[width=8cm]{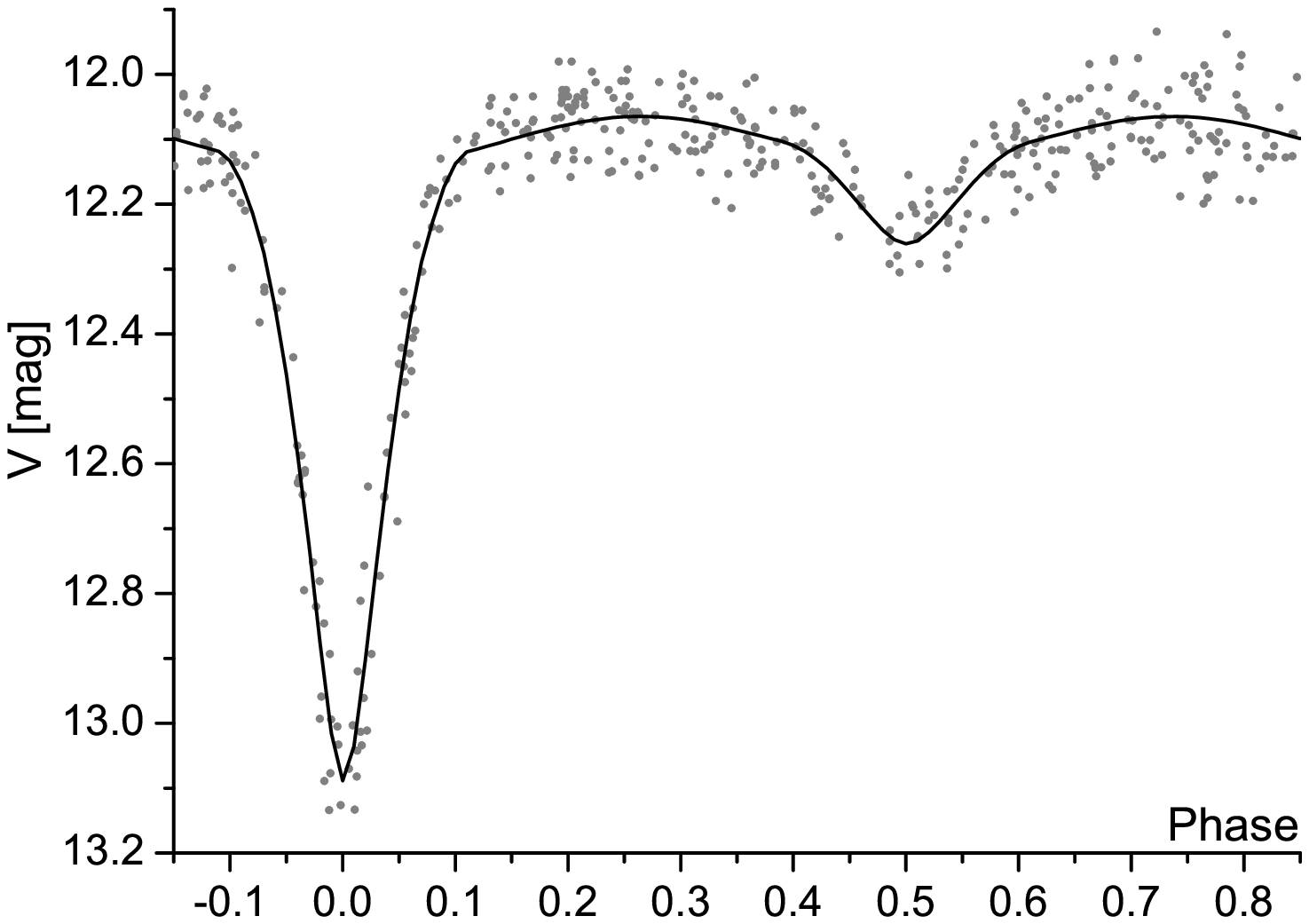}&\includegraphics[width=8cm]{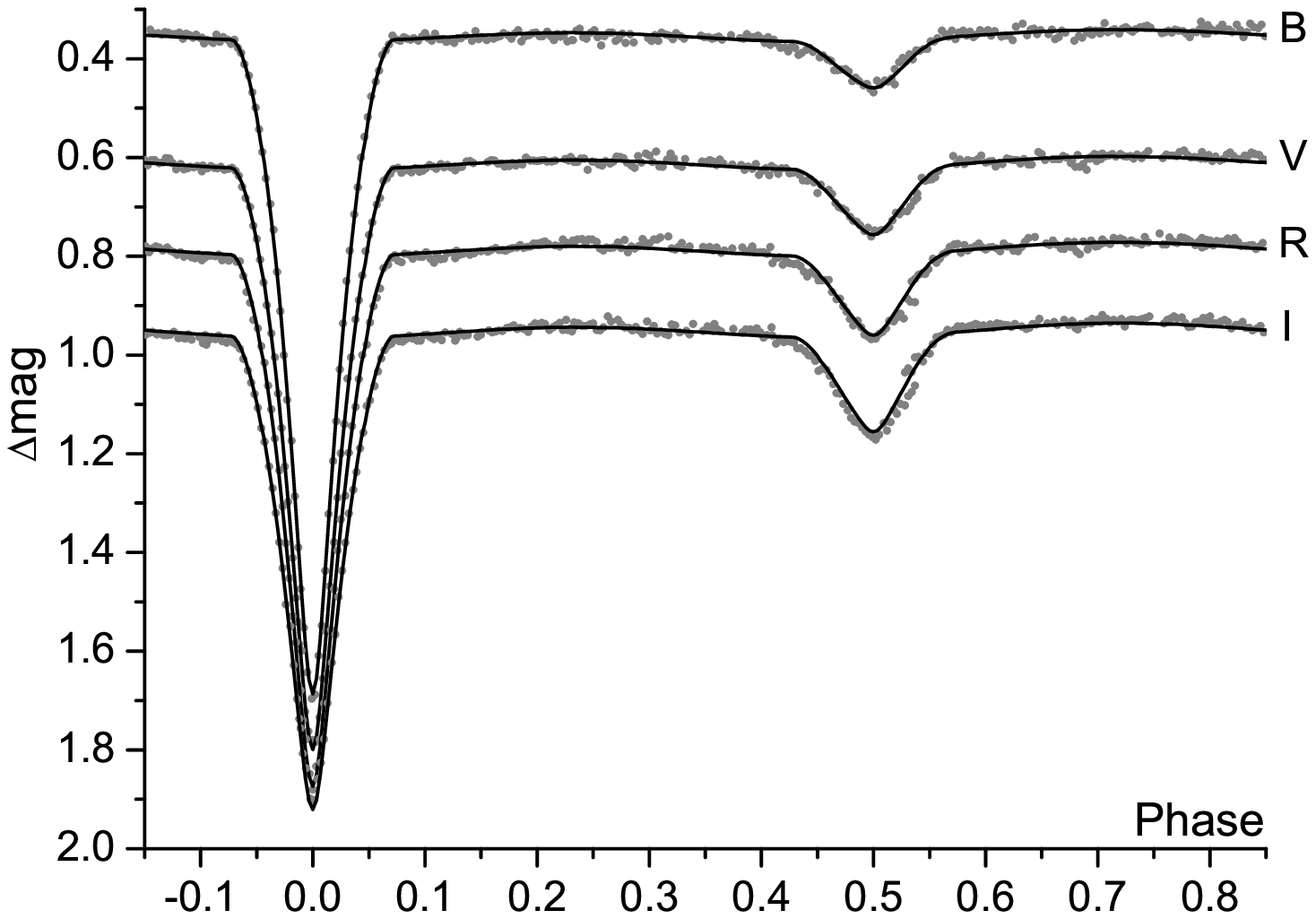}
\end{tabular}
\label{fig2} \caption{The same as Fig. 1 but for FG Gem (left) and UU Leo (right).}

\vspace{5mm}

\begin{tabular}{cc}
\includegraphics[width=8cm]{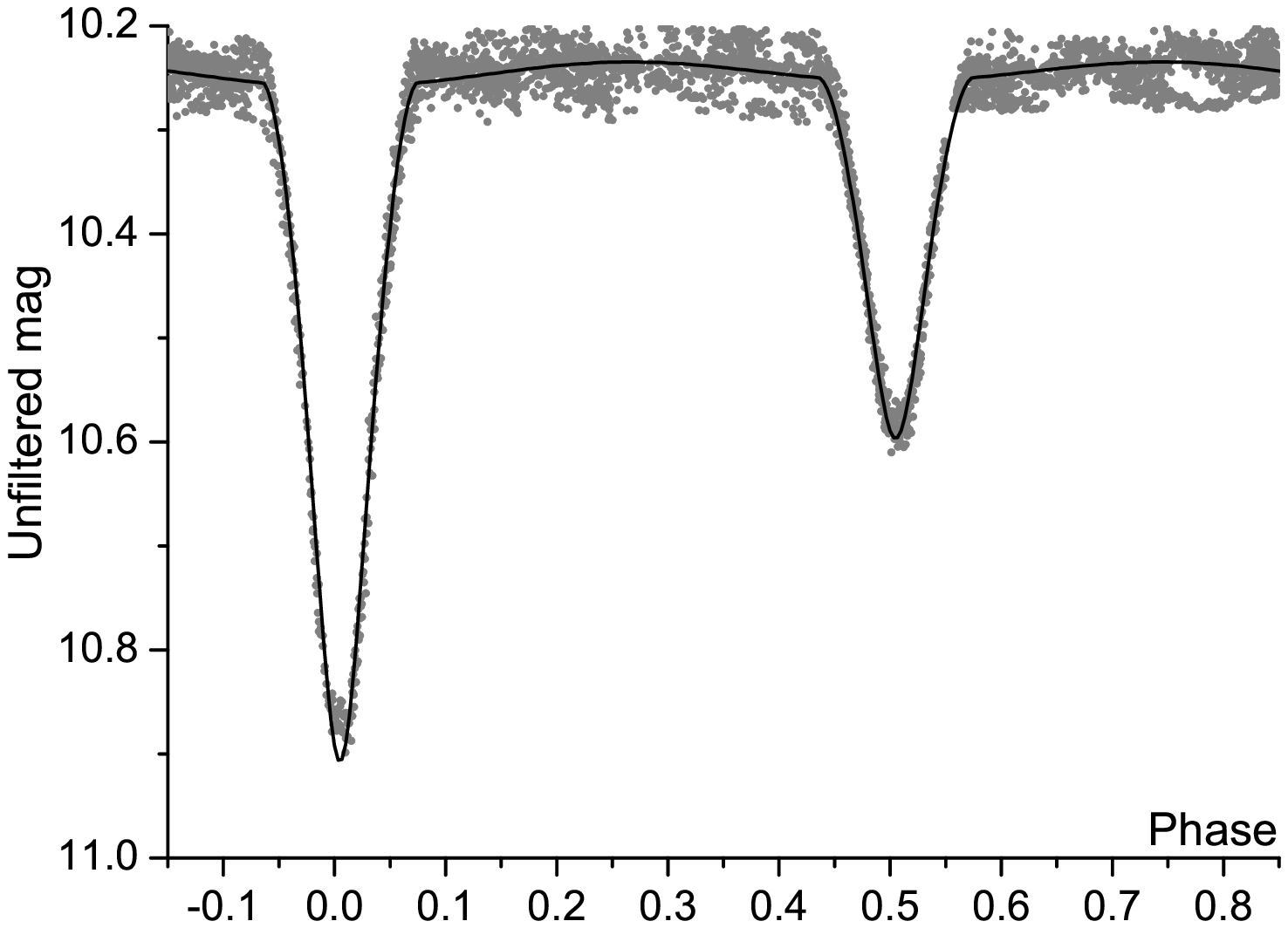}&\includegraphics[width=8cm]{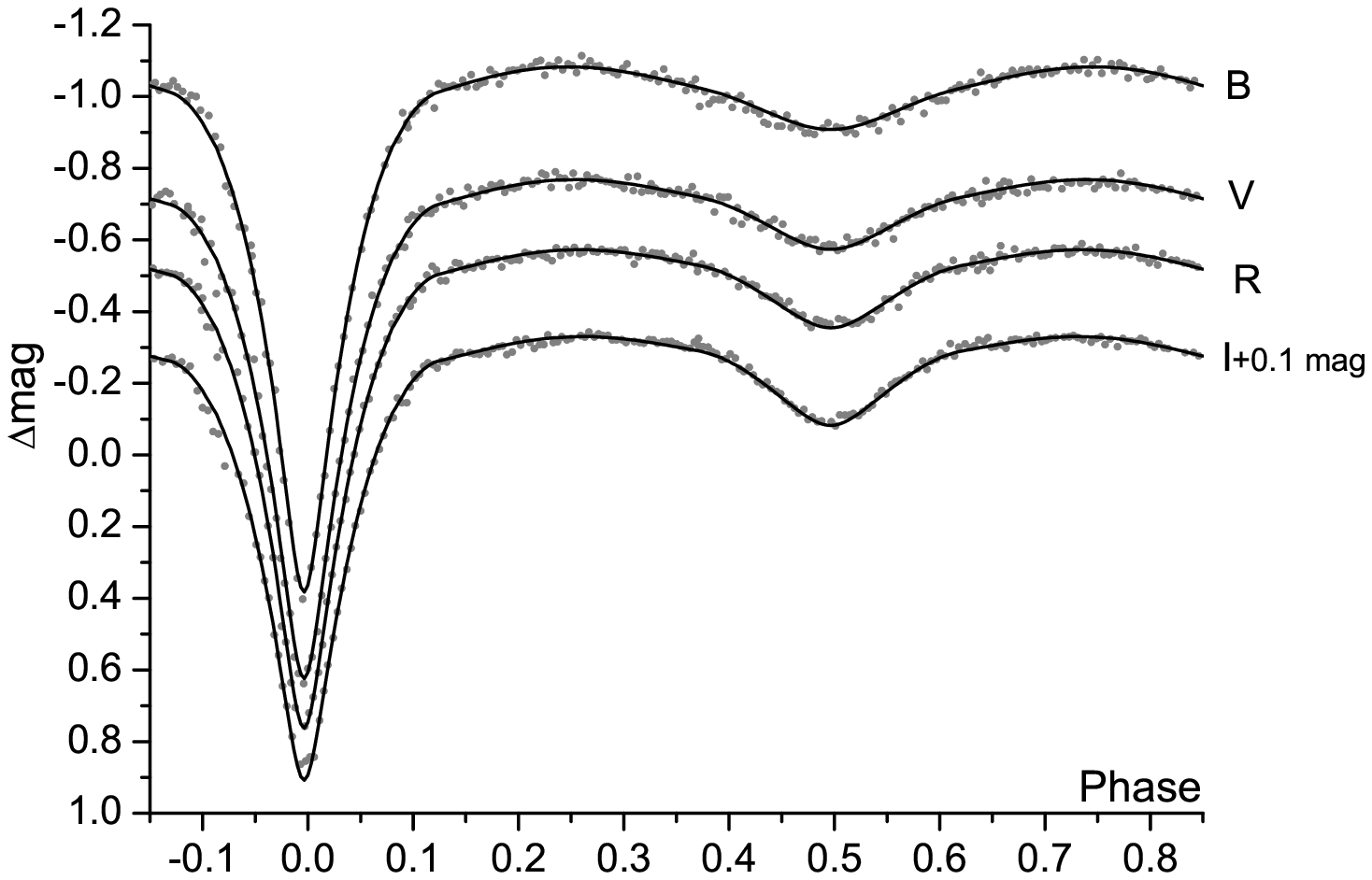}
\end{tabular}
\label{fig3} \caption{The same as Fig. 1 but for CF Tau (left) and AW Vul (right).}
\end{figure}



\begin{table}
\centering

\caption{The light curve solution parameters.}

\label{tab2}


\begin{tabular}{lcccccc}
\hline
\textbf{Parameter}      &\textbf{DP Cep}&\textbf{AL Gem}&\textbf{FG Gem}&\textbf{UU Leo}&\textbf{CF Tau}& \textbf{AW Vul}   \\
\hline
Mode                    &       SD      &       D       &       SD      &       D       &       D       &        SD         \\
i [deg]                 &   85.2 (2)    &   83.0 (1)    &       84 (1)  &   86.8 (1)    &   84.2 (1)    &   86.2 (4)        \\
q (m$_{2}$/m$_{1}$)     &   0.65 (2)    &   0.189 (1)   &   0.41 (3)    &   0.286 (2)   &   0.940 (4)   &   0.55 (1)        \\
$T_1$* [K]              &   7300        &   6650        &   7700        &   9000        &       5950    &       7300        \\
$T_2$ [K]               &   5049 (42)   &   4080 (10)   &   5068 (88)   &   5688 (10)   &   5308 (5)    &   4394 (21)       \\
A$_1$*                  &       1       &       0.5     &       1       &       1       &       0.5     &       1           \\
A$_2$*                  &       0.5     &       0.5     &       0.5     &       0.5     &       0.5     &       0.5         \\
g$_1$*                  &       1       &   0.32        &       1       &       1       &       0.32    &       1           \\
g$_2$*                  &       0.32    &   0.32        &   0.32        &   0.32        &       0.32    &       0.32        \\
$\Omega_{1}$            &   3.87 (3)    &   3.77 (1)    &   3.49 (12)   &   4.98 (2)    &   5.84 (1)    &   3.36 (2)        \\
$\Omega_{2}$            &       3.16    &   2.35 (3)    &   2.69        &   2.65 (1)    &   5.22 (1)    &       2.96        \\
x$_{1,B}$               &       --      &   0.633       &       --      &   0.535       &       --      &       0.591       \\
x$_{2,B}$               &       --      &   0.968       &       --      &   0.754       &       --      &       0.958       \\
x$_{1,V}$               &       0.498   &   0.516       &   0.476       &   0.46        &   **0.579     &       0.496       \\
x$_{2,V}$               &       0.712   &   0.828       &   0.711       &   0.613       &   **0.668     &       0.809       \\
x$_{1,R}$               &       --      &   0.438       &       --      &   0.381       &       --      &       0.417       \\
x$_{2,R}$               &       --      &   0.737       &       --      &   0.527       &       --      &       0.704       \\
x$_{1,I}$               &       --      &   0.362       &       --      &   0.301       &       --      &       0.338       \\
x$_{2,I}$               &       --      &   0.604       &       --      &   0.444       &       --      &       0.583       \\
(L$_1$/L$_T$)$_B$       &       --      &   0.952 (2)   &       --      &   0.794 (3)   &       --      &       0.923 (5)   \\
(L$_2$/L$_T$)$_B$       &       --      &   0.023 (1)   &       --      &   0.103 (1)   &       --      &   0.039 (1)       \\
(L$_3$/L$_T$)$_B$       &       --      &   0.025 (1)   &       --      &   0.104 (2)   &       --      &   0.038 (4)       \\
(L$_1$/L$_T$)$_V$       &   0.59 (2)    &   0.850 (2)   &   0.83 (2)    &   0.747 (3)   &  **0.560 (2)  &   0.902 (6)       \\
(L$_2$/L$_T$)$_V$       &   0.36 (5)    &   0.036 (1)   &   0.12 (2)    &   0.144 (1)   &  **0.397 (1)  &   0.066 (1)       \\
(L$_3$/L$_T$)$_V$       &   0.05 (1)    &   0.113 (2)   &   0.05 (2)    &   0.109 (2)   &  **0.043 (2)  &   0.031 (5)       \\
(L$_1$/L$_T$)$_R$       &       --      &   0.798 (2)   &       --      &   0.712 (3)   &       --      &   0.884 (7)       \\
(L$_2$/L$_T$)$_R$       &       --      &   0.049 (1)   &       --      &   0.173 (4)   &       --      &   0.092 (1)       \\
(L$_3$/L$_T$)$_R$       &       --      &   0.153 (1)   &       --      &   0.115 (3)   &       --      &   0.024 (5)       \\
(L$_1$/L$_T$)$_I$       &       --      &   0.728 (2)   &       --      &   0.666 (3)   &       --      &   0.849 (8)       \\
(L$_2$/L$_T$)$_I$       &       --      &   0.064 (1)   &       --      &   0.204 (1)   &       --      &   0.123 (1)       \\
(L$_3$/L$_T$)$_I$       &       --      &   0.208 (1)   &       --      &   0.130 (3)   &       --      &   0.028 (6)       \\
\hline
$<$L$_3$/L$_T$$>$ (\%)  &   5 (1)       &   12.5 (1)    &   5 (2)       &   11.5 (3)    &   4.3 (2)     &   3.0 (5)         \\
\hline
                                        \multicolumn{7}{c}{\emph{Fractional radii}}                                         \\
\hline
$r_1$ (pole)            &   0.308       &   0.279       &   0.322       &   0.208       &   0.203       &       0.351       \\
$r_1$ (point)           &   0.332       &   0.285       &   0.341       &   0.210       &   0.208       &       0.390       \\
$r_1$ (side)            &   0.316       &   0.283       &   0.330       &   0.209       &   0.205       &       0.364       \\
$r_1$ (back)            &   0.325       &   0.284       &   0.337       &   0.209       &   0.207       &       0.377       \\
$r_2$ (pole)            &   0.321       &   0.196       &   0.284       &   0.220       &   0.223       &       0.307       \\
$r_2$ (point)           &   0.456       &   0.221       &   0.409       &   0.248       &   0.231       &       0.438       \\
$r_2$ (side)            &   0.336       &   0.201       &   0.296       &   0.225       &   0.223       &       0.320       \\
$r_2$ (back)            &   0.368       &   0.214       &   0.329       &   0.240       &   0.223       &       0.353       \\

\hline
                                  \multicolumn{7}{c}{\emph{Spot parameters}}                                                \\
\hline
Lat [deg]               &        --     &   90 (2)      &       --      &   90 (2)      &       --      &           --      \\
Long [deg]              &       --      &   234 (6)     &       --      &   28 (3)      &       --      &           --      \\
R [deg]                 &       --      &   21 (1)      &       --      &   24 (2)      &       --      &           --      \\
T$_f$                   &       --      &   0.89 (5)    &       --      &   0.88 (3)    &       --      &           --      \\
\hline
$\sum res^{2}$          &       0.953   &    0.085      &     0.745     &    0.153      &      0.806    &           0.099   \\
\hline
\multicolumn{7}{l}{\textit{*assumed}, \textit{**Bolometric values}, \textit{L$_T$=L$_1$+L$_2$+L$_3$}, \textit{D=Detached}, \textit{SD=SemiDetached}}  \\
\end{tabular}
\end{table}

\section{Search for pulsations}
\label{4}

Two of the systems of the present study, namely UU Leo and AW Vul, were listed as candidates for containing a $\delta$ Sct component \citep{b23}. In that work the candidacy criterion was the EB's spectral type (A-F). The rest of systems, except CF Tau, are also inside this range but they were lacking from that list. Consequently, we searched for short-period pulsations in the data of these five systems. For this, the theoretical LCs of all filters were subtracted from the respective observed ones and a frequency search in the residuals between 3 to 80 c/d \citep[typical for $\delta$~Sct stars]{b44,b23} was performed. The programme \emph{PERIOD04} v.1.2 \citep{b45} that is based on classical Fourier analysis was used. Finally, the results showed that neither star demonstrated convincing pulsational behaviour in the selected range of frequencies with a signal-to-noise ratio (S/N) higher than 4.

\section{Absolute parameters \& Evolutionary stage}
\label{5}

Although no radial velocity curves exist for the systems, we can make fair estimates of their absolute parameters. These are listed in Table \ref{tab3}. The mass of the primary component of each system was assumed from its spectral type, while masses of the secondaries followed from the adopted mass ratios given above (see Table \ref{tab2}). The semi-major axes (a), used to calculate mean radii, then follow from Kepler's law. Formal errors are
indicated in parentheses alongside adopted values. In Fig. 4, the location of both components of each system in the Mass-Radius (M -- R) diagram is illustrated.

The primary components of FG Gem, UU Leo and AW Vul are inside the MS band, while the respective ones of DP Cep and AL Gem are located at the TAMS limit. Contrary to that, the primary of CF Tau, and all systems' secondaries were found to be evolved stars lying in the range 3.64$<$logg$<$3.95, which corresponds to the subgiant luminosity classes.

A rough distance (D) estimation of each system, based on the photometric parallax method, was performed. The calculated absolute magnitudes of the systems, $M_{bol,system}$, were calculated by using the luminosities of the components ($L_1$ and $L_2$) and then transformed to standard ones $M_{V,system}$. The apparent V-magnitudes were taken from literature sources. In particular, in the ASAS 5$^{th}$ catalogue \citep{b17} the binarity was taken into account and the combined, out of eclipses, apparent magnitudes are given for the systems AL Gem (V=9.77 mag), FG Gem (V=12.04 mag), CF Tau (V=10.05 mag) and AW Vul (V=11.11 mag). For DP Cep there are not many magnitude estimations, therefore the most probable value comes from \emph{The Guide Star Catalogue} Version 2.3.2 \citep{b54} as V=12.9 mag. Contrarily, for UU Leo a lot of information in V-passband is available, and it was very difficult to select a unique value. Thus, a value of of V=11.74 mag (average of V-apparent magnitudes given in several catalogues) was settled.

\begin{table}[t]
\centering

\caption{Physical and orbital parameters of the components of the binaries and distance estimations of the systems.}

\label{tab3}

\begin{tabular}{l cc cc cc}
\hline
\textbf{Parameter}      &\textbf{DP Cep}&\textbf{AL Gem}&\textbf{FG Gem}&\textbf{UU Leo}&\textbf{CF Tau}& \textbf{AW Vul}   \\
\hline
M$_1$* [M$_\odot$]      &       1.6     &       1.4     &       1.75    &       2.5     &       1.05    &   1.6             \\
M$_2$** [M$_\odot$]     &   1.05 (3)    &   0.26 (1)    &   0.71 (6)    &   0.72 (1)    &   0.99 (1)    &   0.87 (1)        \\
R$_1$ [R$_\odot$]       &   2.2 (1)     &   1.79 (4)    &   1.7 (2)     &   1.92 (5)    &   2.20 (5)    &   1.84 (8)        \\
R$_2$ [R$_\odot$]       &   2.4 (1)     &   1.29 (3)    &   1.5 (2)     &   2.02 (5)    &   2.42 (5)    &   1.65 (7)        \\
T$_1$* [K]              &   7300        &   6650        &   7700        &   9000        &    5950       &   7300            \\
T$_2$ [K]               &   5049 (42)   &   4080 (10)   &   5068 (88)   &   5688 (10)   &    5308 (5)   &   4394 (21)       \\
Log g$_1$ [cm/s$^2$]    &   3.95 (5)    &   4.08 (2)    &   4.23 (9)    &   4.29 (3)    &   3.77 (2)    &   4.11 (4)        \\
Log g$_2$ [cm/s$^2$]    &   3.70 (5)    &   3.64 (2)    &   3.92 (9)    &   3.67 (3)    &   3.66 (2)    &   3.95 (4)        \\
L$_1$ [L$_\odot$]       &   12 (1)      &   5.6 (3)     &   9 (2)       &   22 (1)      &   5.4 (2)     &   8.6 (8)         \\
L$_2$ [L$_\odot$]       &   3.3 (4)     &   0.42 (3)    &   1.4 (3)     &   3.8 (2)     &   4.2 (2)     &   0.9 (1)         \\
M$_{bol, 1}$ [mag]      &   2 (1)       &   2.9 (5)     &   2 (2)       &   1.4 (6)     &   2.9 (5)     &   2.4 (9)         \\
M$_{bol, 2}$ [mag]      &   3 (2)       &   5.7 (3)     &   4 (2)       &   3.3 (6)     &   3.2 (6)     &   4.7 (9)         \\
a$_1$ [R$_\odot$]       &   2.8 (4)     &   1.0 (1)     &   1.5 (5)     &   2.0 (2)     &   5.2 (2)     &   1.8 (2)         \\
a$_2$ [R$_\odot$]       &   4.22 (3)    &   5.35 (4)    &   3.6 (1)     &   7.0 (1)     &   5.5 (1)     &   3.3 (1)         \\
D [pc]                  &  1781 (848)   &  247 (77)     & 1091 (1487)   & 1221 (429)    &  353 (170)    &   579 (338)       \\
\hline
\multicolumn{7}{l}{\emph{*assumed, **M$_2$=q$_{ph}\cdot$M$_1$}}                                                             \\
\end{tabular}
\end{table}

\begin{figure}[h!]
\centering

\includegraphics[width=8cm]{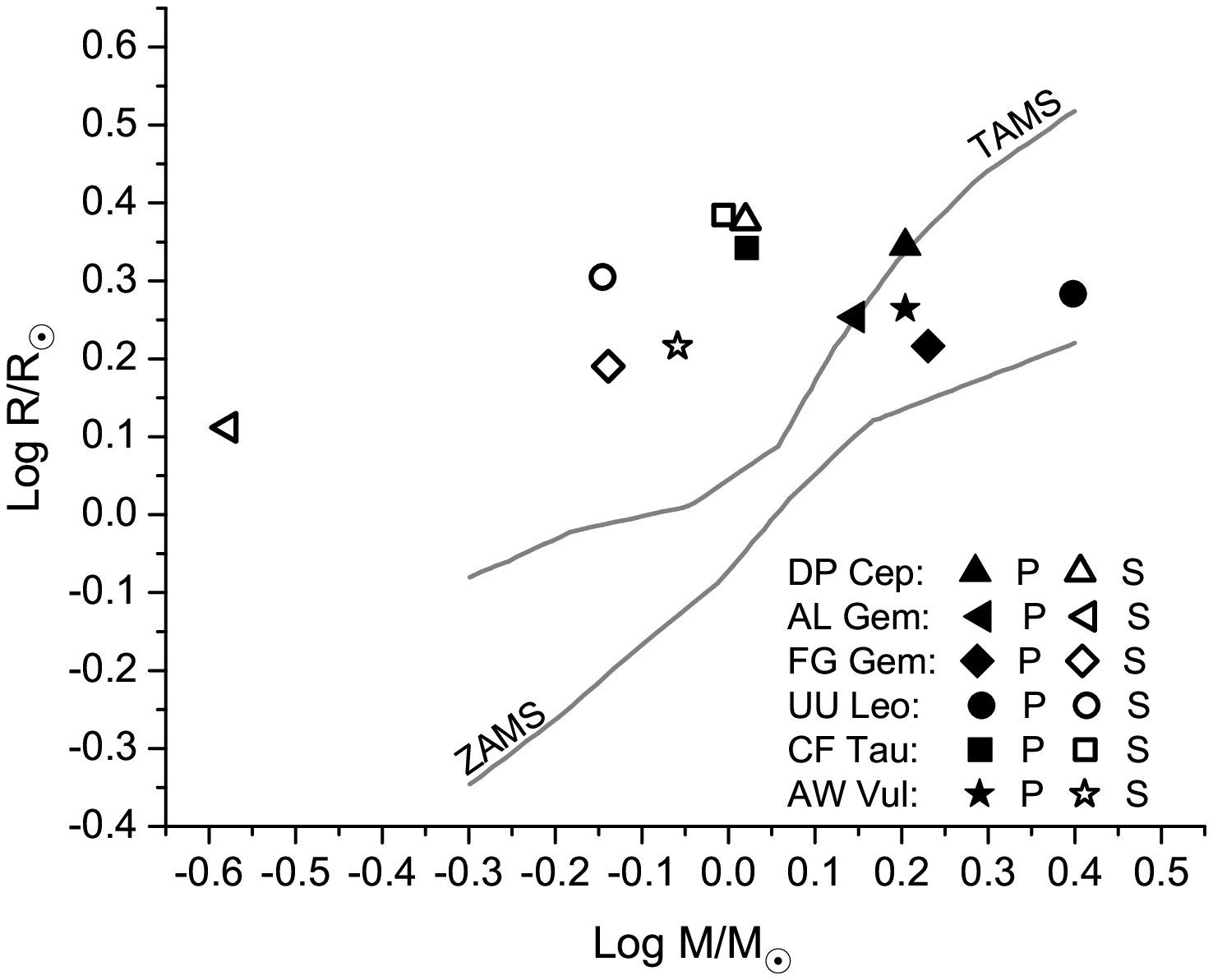}
\label{fig4} \caption{The location of the components (open and filled symbols) of the systems in the M -- R diagram. The indicators $P$ and $S$ refer to the primary and secondary components, respectively. The lines shown are solar-metalicity Terminal Age Main Sequence (TAMS) and Zero Age Main Sequence (ZAMS) \citep{b57}.}
\end{figure}

\section{O -- C diagram analysis}
\label{6}

The computation of the parameters of the third-body orbit is a classical inverse problem with five parameters to be found, namely the period of the third body $P_3$, the HJD of the periastron passage $T_0$, the semi-amplitude $A$ of the $Light~Time~Effect$ (hereafter LITE), the argument of periastron $\omega$ and the eccentricity of the third body $e_3$. The ephemeris of the system ($JD_0$ and $P$) has to be calculated together with the parameters of LITE. The mass function of the third body and its minimal mass $M_{3,min}$ (for coplanar orbit) could be computed from this set of parameters. If the distance of the system is determined (see Table \ref{tab3}), one is able to calculate the tertiary component's predicted angular separation $a_3$ from the EB (semi-major axis of the orbit as projected on the sky), while given the luminosity fraction of each component of the triple system (see Table \ref{tab2}) the magnitude difference between the third component and the eclipsing pair $\Delta$M can be also computed. The $a_3$ and $\Delta$M values are given here for a prospective interferometric detection of the third bodies.

The O -- C diagrams were analysed using the least squares method in a Matlab-code designed by P.Z. \citep{b46}. The weights assigned to individual observations were set at $w=1$ for visual observations, 5 for photographic and 10 for CCD and photoelectric observations. In Figs. 5-7 full circles represent times of primary minima and open circles those of the secondary minima, where the bigger the symbol, the bigger the weight assigned. Each figure
consists of two panels (upper and lower). The upper one shows the fit on the O -- C data, while in the lower one the distribution of the residuals after the subtraction of the adopted solution is illustrated. In Table \ref{tab5} the parameters derived from our analysis together with the minimal masses of the additional bodies orbiting the EBs and the new ephemerides of the systems are given.

\subsection{DP Cep}

The O -- C analysis of this system was based on 78 times of minima covering $\sim$60 yrs. 66 of them were taken from the literature (44 visual, 12 photographic and 10 CCD), and the rest 12 were derived, by using the \citet{b58} method, from INTEGRAL/OMC \citep{b31} and NSVS (\emph{Northern Sky Variability Survey}, \citealt{b59}) projects data (see Table \ref{tab4}). Initially, the following ephemeris of \citet{b47} was used to create the O -- C diagram of DP Cep:
\begin{equation}
{\rm Min.I} = {\rm HJD}~2433622.268+1.26996323^{d}\times~E
\end{equation}
A theoretical LITE curve was selected for fitting the times of minima, due to their periodic distribution.

\begin{table}
\centering

\caption{The observed times of minima for DP Cep from the INTEGRAL/OMC and NSVS data.}

\label{tab4}
\vspace{1mm} 

\begin{tabular}{cccc}
\hline
\textbf{HJD (2400000.0+)} &\textbf{Filter}&	\textbf{Type} & \textbf{Source}  \\
\hline
51356.025 (3)	          &	      --	  &	      I	      &  NSVS            \\
51484.290 (2)	          &	      --	  &	      I	      &  NSVS            \\
51484.926 (2)	          &	      --	  &	      II	  &  NSVS            \\
52997.453 (2)	          &	      V	      &	      II	  & INTEGRAL/OMC     \\
52998.082 (1)	          &	      V	      &	      I	      & INTEGRAL/OMC     \\
53357.484 (1)	          &	      V	      &	      I	      & INTEGRAL/OMC     \\
53358.124 (2)	          &	      V	      &	      II	  & INTEGRAL/OMC     \\
53716.889 (1)	          &	      V	      &	      I	      & INTEGRAL/OMC     \\
53720.699 (1)	          &	      V	      &	      I	      & INTEGRAL/OMC     \\
53732.129 (1)	          &	      V	      &	      I	      & INTEGRAL/OMC     \\
53741.016 (3)	          &	      V	      &	      I	      & INTEGRAL/OMC     \\
53888.335 (2)	          &	      V	      &	      I	      & INTEGRAL/OMC     \\
\hline
\end{tabular}
\end{table}

\subsection{AL Gem}

The O -- C diagram for AL Gem was initially constructed using the ephemeris \citep{b47}:
\begin{equation}
{\rm Min.I} = {\rm HJD}~2426324.4424+1.39134001^{d}\times~E
\end{equation}
and consists of 94 literature times of minima (48 visual, 18 photographic, 3 photoelectric and 25 CCD). These data cover an interval from 1910 to the present. In view of apparent low amplitude cyclic variations, we included a LITE curve fitter.

The solution seems to describe very well the O -- C points in total. However, the residuals of the most recent data (after 2005) do not follow exactly the theoretical curve, indicating that another mechanism might implicate the period changes, thus, new minima timings are needed in the next ten years in order to verify its existence.

\subsection{FG Gem}

For the O -- C diagram of FG Gem 88 times of minima taken from the literature (49 visual, 13 photographic, and 26 CCD) were used. The times of minima come from 1930 up to 2010, with a long gap between 1950-1975. Initially, the following ephemeris \citep{b47}:
\begin{equation}
{\rm Min.I} = {\rm HJD}~2427102.4135+0.81912814^{d}\times~E
\end{equation}
was used to create the diagram. LITE curve, indicated by the periodic modulation in the O -- C, was selected as the fitting function. Due to the poor coverage near periastron of the orbit, its high eccentricity is still questionable.

\subsection{UU Leo}

The current O -- C diagram of UU Leo includes 99 times of minima taken from the literature since 1910 (57 visual, 12 photographic, and 30 CCD observations). The following ephemeris \citep{b47} was used to compute, initially, the O -- C points from all the compiled data:
\begin{equation}
{\rm Min.I} = {\rm HJD}~2445397.472+1.67974627^{d}\times~E
\end{equation}
A LITE curve, associated with the periodic trend of the O -- C points, characterized the fitting function. In this case the residuals from 2000 to date suggest that an additional mechanism occurs in the system and play a role to its period changes. Their distribution cannot be yet characterized as strictly periodic, but certainly more times of minima in the next 5-10 years will give a clear answer.

\subsection{CF Tau}

58 timings of minima from published papers (20 visual, 4 photographic, and 34 CCD) covering the time span 1930-2010, and the following ephemeris \citep{b47}:
\begin{equation}
{\rm Min.I} = {\rm HJD}~2430651.242+2.7558772^{d}\times~E
\end{equation}
were used to create the O -- C diagram of CF Tau. Similarly to the previous cases, the LITE was adopted as the most plausible mechanism for describing the period variations. The results indicated rather high eccentricity, similarly to FG Gem, but this is fairly well proved by the significant number of minima observations during the periastron of the long orbit.

\subsection{AW Vul}

This system has the largest number of available minima timings in comparison with the other ones. Particularly, 253 times of minima (208 visual, 15 photographic, and 30 CCD observations) were collected from the literature and according to the ephemeris given by \citet{b47}:
\begin{equation}
{\rm Min.I} = {\rm HJD}~2446285.4653+0.80645138^{d}\times~E
\end{equation}
the O -- C diagram of the system was constructed, and a LITE function was selected to fit it. The residuals did not present any obvious variable behaviour, except the ones of the last decade. However, since the current fit is based mostly on visual minima, it is probable that in the following years, and after new timings of minima determination, a more accurate solution will be achieved.


\begin{figure}[h!]
\centering
\begin{tabular}{cc}
\includegraphics[width=8cm]{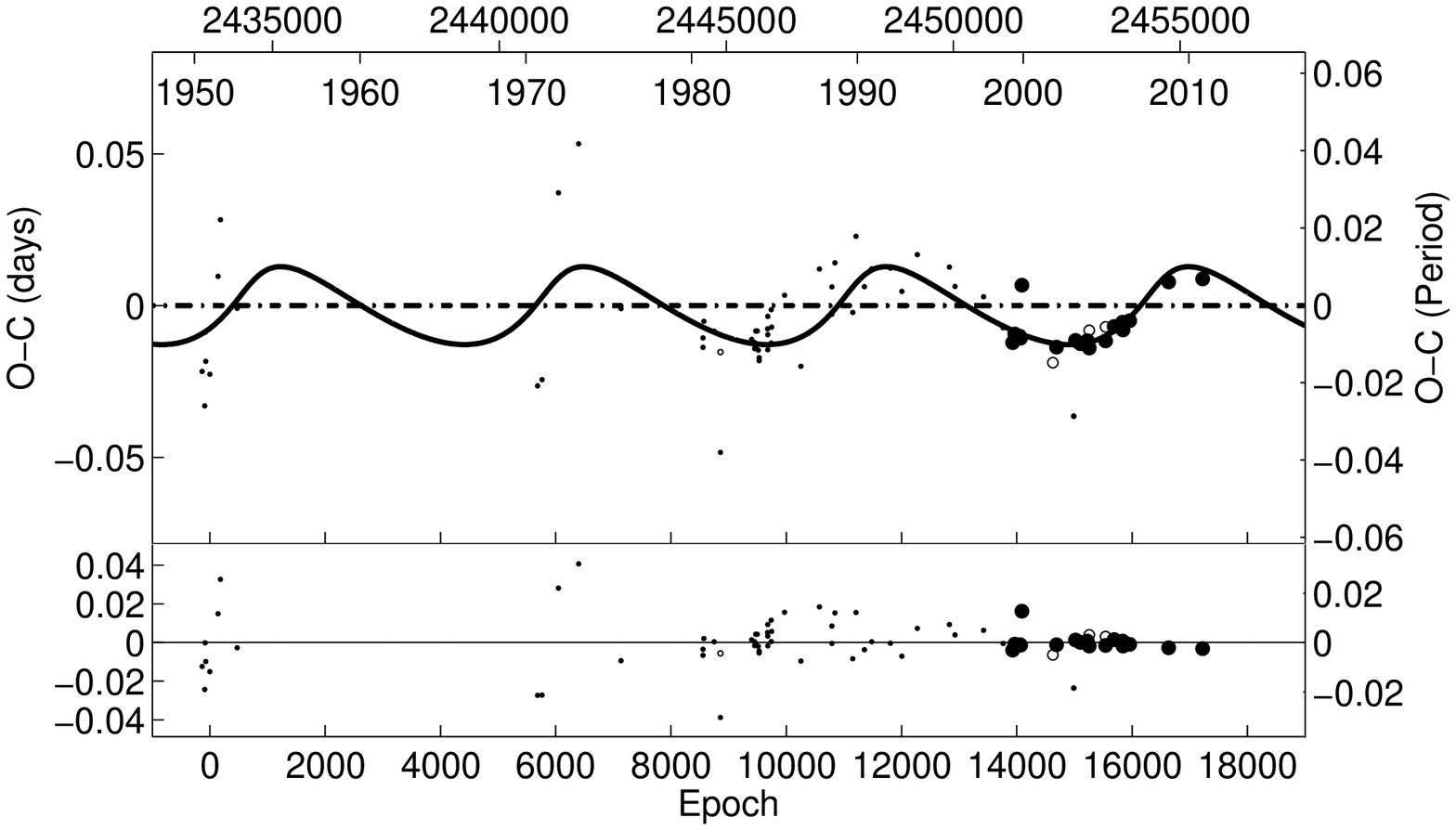}&\includegraphics[width=8cm]{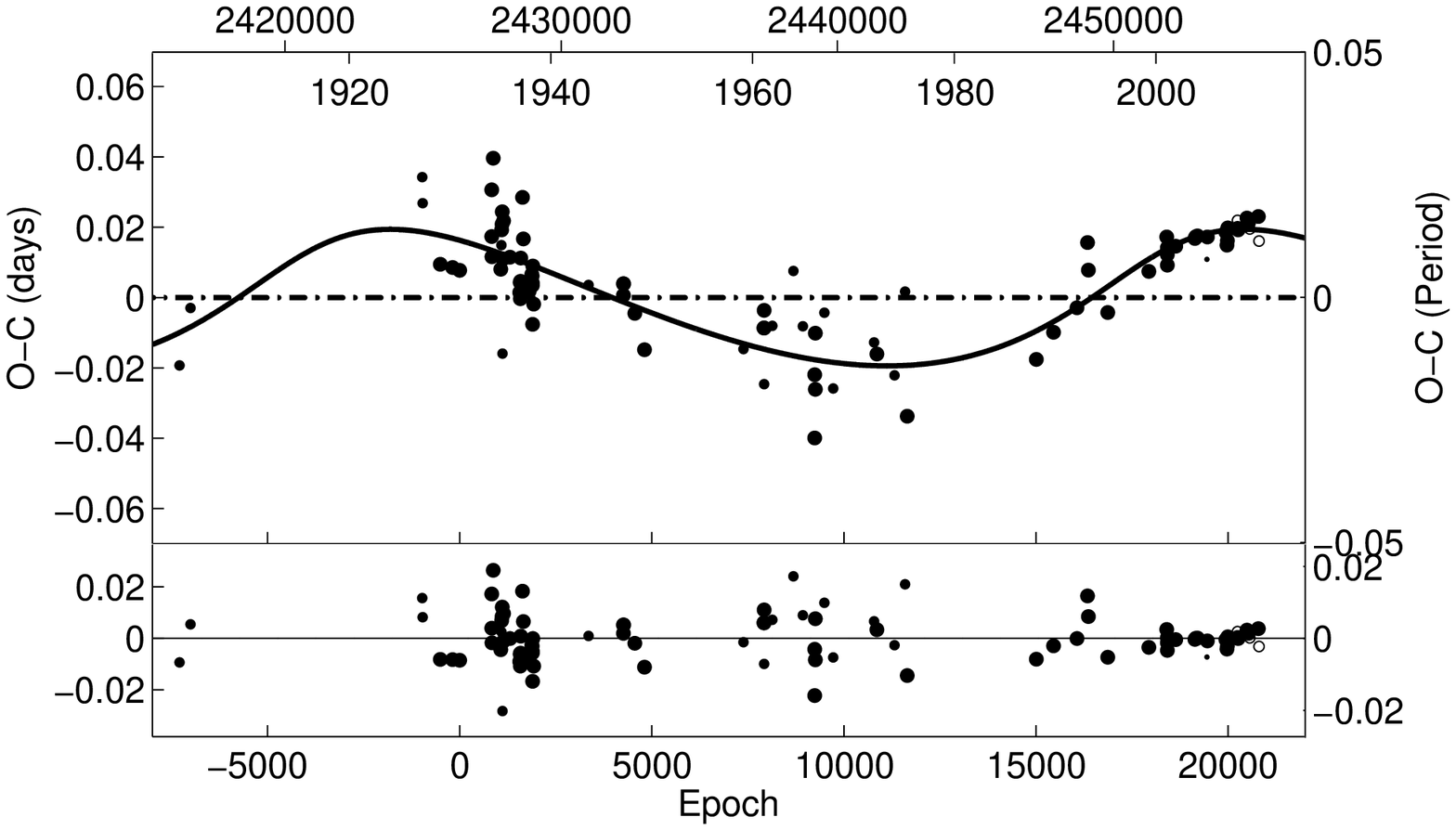}
\end{tabular}
\label{fig5} \caption{The O -- C diagrams of DP Cep (left) and AL
Gem (right) fitted by a LITE curve (upper part) and the residuals
after the subtraction of the solution (lower part).}

\vspace{5mm}
\begin{tabular}{cc}
\includegraphics[width=8cm]{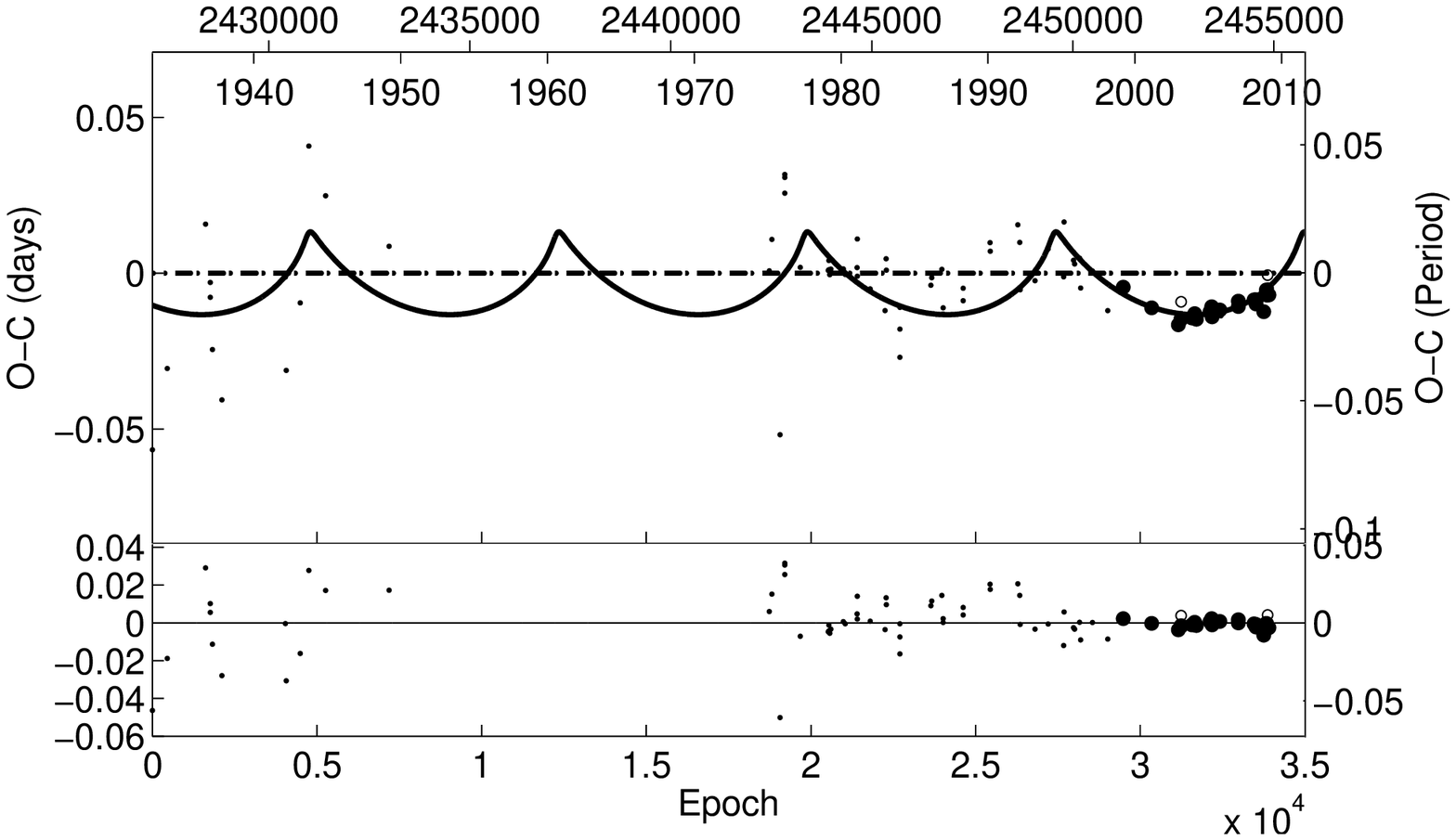}&\includegraphics[width=8cm]{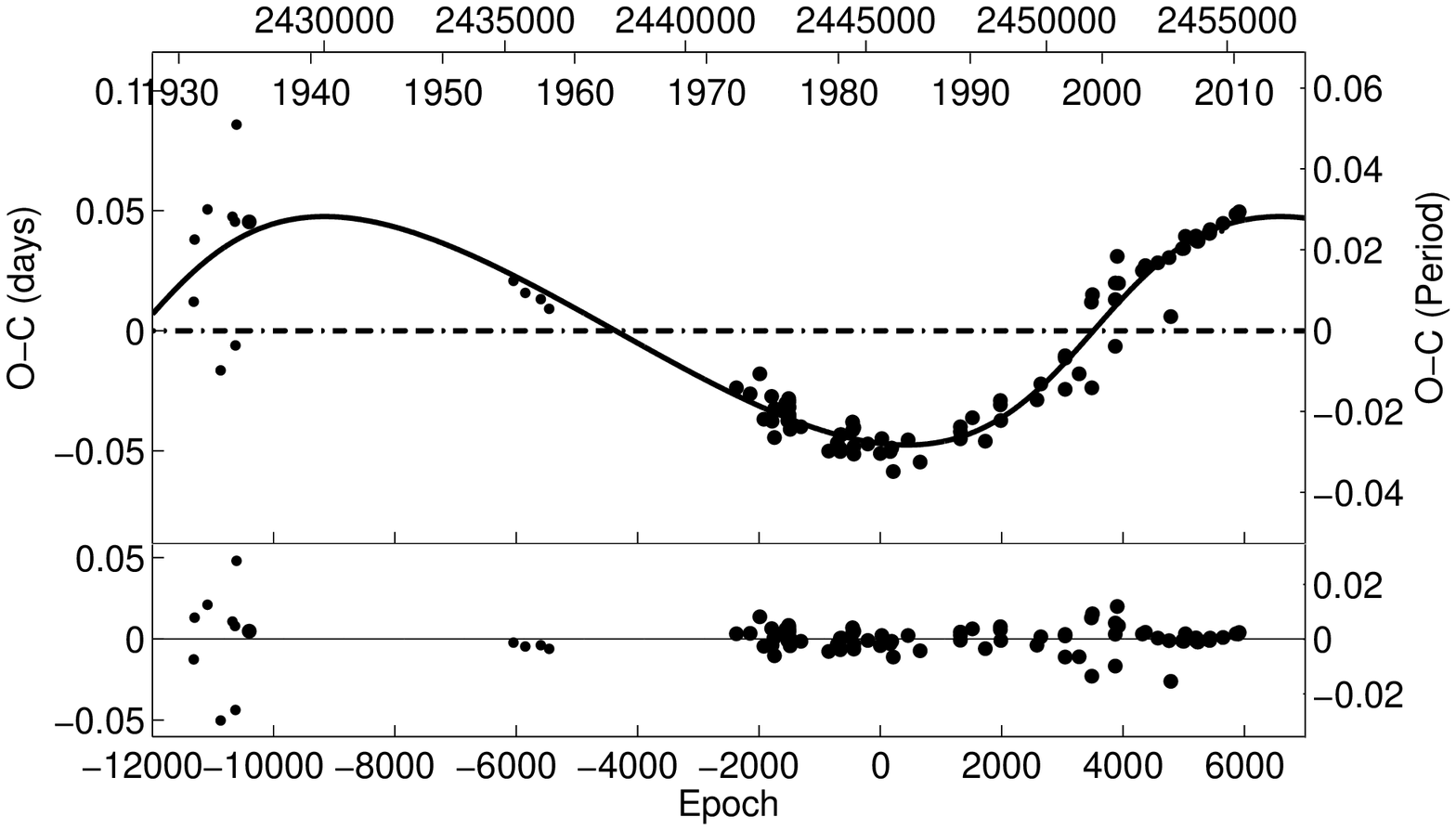}
\end{tabular}
\label{fig6} \caption{The same as Fig. 5 but for FG Gem (left) and
UU Leo (right).}

\vspace{5mm}
\begin{tabular}{cc}
\includegraphics[width=8cm]{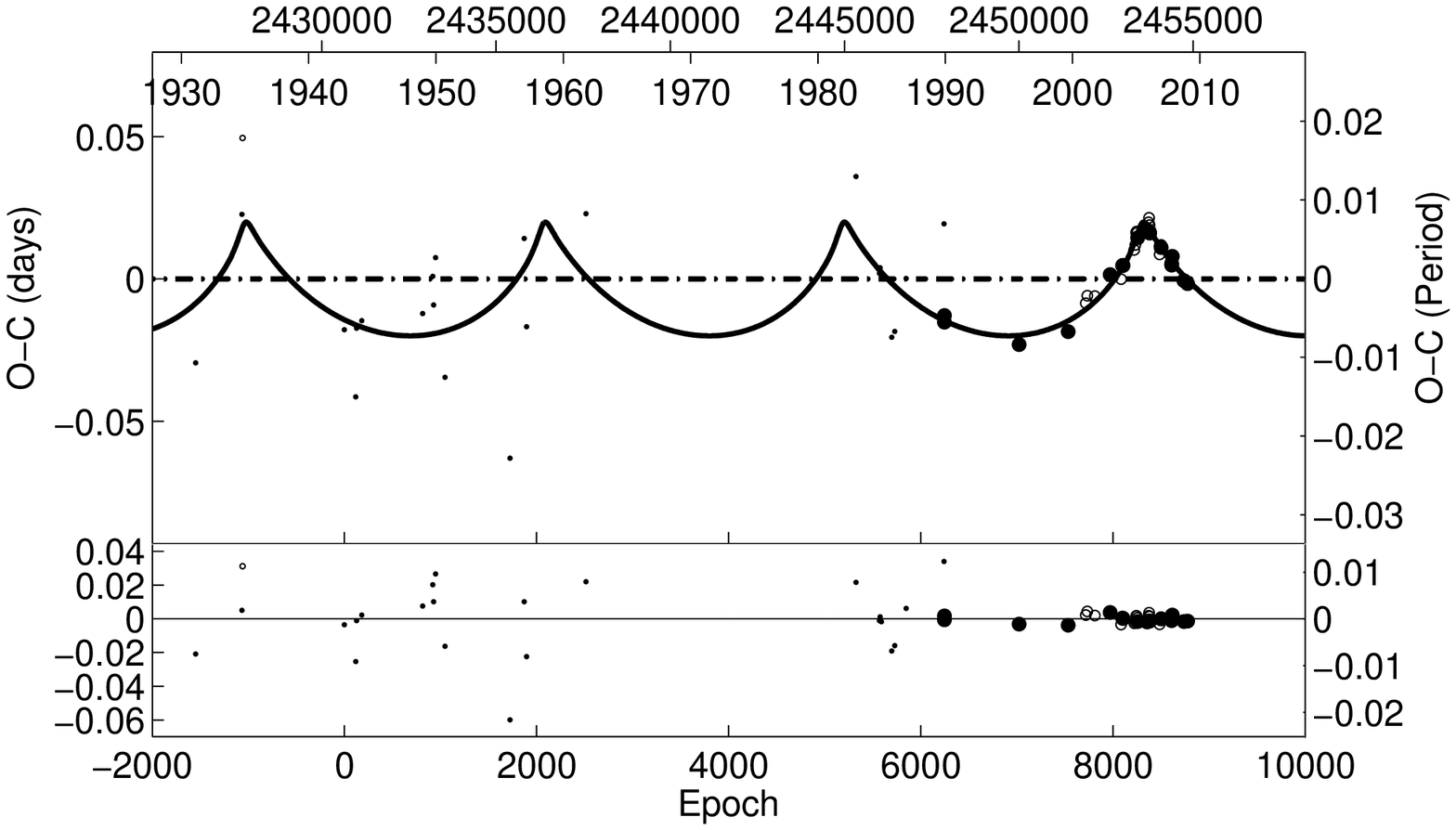}&\includegraphics[width=8cm]{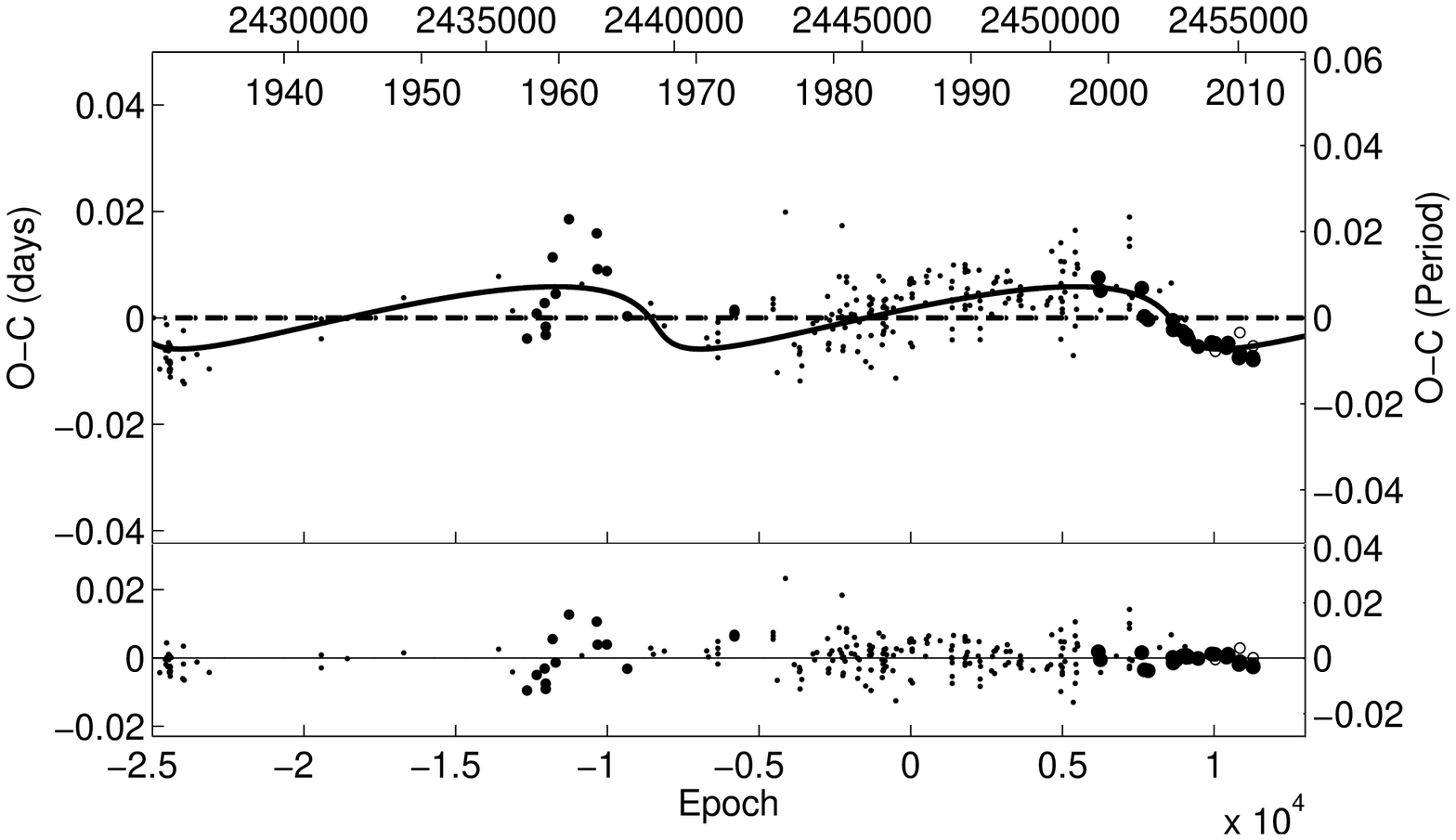}
\end{tabular}
\label{fig7} \caption{The same as Fig. 5 but for CF Tau (left) and
AW Vul (right).}
\end{figure}

\begin{table}[t]
\centering{

\caption{The results of the O -- C analysis.}

\label{tab5}

\vspace{1mm} 

\begin{tabular}{lccc}
\hline \textbf{Parameter}   &   \textbf{DP Cep} &   \textbf{ALGem}  &    \textbf{FG Gem}    \\
\hline
                                \multicolumn{4}{c}{\emph{Eclipsing Binary}}                 \\
\hline
JD$_0$ [HJD]                &   2433622.279 (5) &   2426324.432 (4) &   2427102.427 (4)     \\
P [days]                    &   1.2699625 (4)   &   1.3913414 (3)   &   0.8191276 (2)       \\
M$_1$+M$_2$ [M$_\odot$]     &       1.6 + 1.04  &   1.4 + 0.26      &   1.75 + 0.71         \\
\hline
                                \multicolumn{4}{c}{\emph{LITE \& Third Body}}               \\
\hline
P$_3$ [yrs]                 &   18.3 (4)        &   85 (6)          &   16.9 (3)            \\
T$_0$ [HJD]                 &   2447861 (613)   &   2451427 (2521)  &   2433650 (315)       \\
A [days]                    &   0.013 (2)       &   0.019 (2)       &   0.013 (2)           \\
$\omega$ [deg]              &   33 (36)         &   35 (37)         &   68 (36)             \\
e$_3$                       &   0.4 (3)         &   0.3 (2)         &   0.8 (2)             \\
f(m$_3$) [M$_\odot$]        &   0.040 (1)       &   0.006 (1)       &   0.052 (7)           \\
M$_{3,min}$ [M$_\odot$]     &   0.77 (1)        &   0.28 (2)        &   0.82 (5)            \\
$a_3$ [mas]                 &       6 (5)       &     98 (42)       &      9 (5)            \\
$\Delta$M [mag]             &       3.2 (2)     &       2.2 (3)     &       3.3 (6)         \\
\hline
$\sum Wres^{2}$             &   0.015           &   0.060           &   0.017               \\
\hline
\\
\hline
\textbf{Parameter}          &   \textbf{UU Leo} &   \textbf{CF Tau} &   \textbf{AW Vul}     \\
\hline
                                \multicolumn{4}{c}{\emph{Eclipsing Binary}}                 \\
\hline
JD$_0$ [HJD]                &   2445397.507 (4) &   2430651.248 (5) &   2446285.463 (1)     \\
P [days]                    &   1.6797524 (8)   &   2.7558776 (8)   &   0.8064510 (1)       \\
M$_1$+M$_2$ [M$_\odot$]     &   2.5 + 0.72      &   1.05 + 0.99     &   1.6 + 0.87          \\
\hline
                                \multicolumn{4}{c}{\emph{LITE \& Third Body}}               \\
\hline
P$_3$ [yrs]                 &   72 (1)          &   23.5 (3)        &   38 (1)              \\
T$_0$ [HJD]                 &   2424839 (862)   &   2453533 (208)   &   2439585 (436)       \\
A [days]                    &   0.048 (2)       &   0.020 (3)       &   0.006 (1)           \\
$\omega$ [deg]              &   0 (11)          &   72 (24)         &   197 (9)             \\
e$_3$                       &   0.3 (1)         &   0.8 (1)         &   0.7 (2)             \\
f(m$_3$) [M$_\odot$]        &   0.129 (1)       &   0.084 (5)       &   0.0019 (1)          \\
M$_{3,min}$ [M$_\odot$]     &   1.40 (9)        &   0.89 (2)        &   0.24 (1)            \\
$a_3$ [mas]                 &       24 (9)      &       33 (12)     &     27 (38)           \\
$\Delta$M [mag]             &       2.3 (1)     &       4.3 (3)     &       3.8 (3)         \\
\hline
$\sum Wres^{2}$             &   0.080           &   0.012           &   0.008               \\
\hline

\end{tabular}}
\end{table}

\section{Discussion and conclusions}
\label{7}

Complete LCs of the eclipsing systems AL Gem, UU Leo and AW Vul were obtained, and together with the ones of DP Cep, FG Gem and CF Tau obtained from sky surveys were analysed in order to derive their geometrical and absolute parameters. In addition, the LC residuals of all systems, except for CF Tau, were tested for possible pulsations but the results were negative. O -- C analysis of each system was performed and showed probable existence of a tertiary component orbiting the common center of mass. In all cases, the contribution of a third light to the total luminosity was taken into account in the LC analysis, and its fraction $L_{3,LC}(\%)$ (see Table \ref{tab2}) was derived.

According to the minimal mass of the third body found from the O -- C analysis, one is able to derive its absolute luminosity by assuming its MS nature and by using the mass-luminosity relation for such stars (L$\sim$M$^{3.5}$); then, the calculation of its light contribution to the total luminosity of the system, given the absolute luminosities of the eclipsing components $L_1$ and $L_2$ (see Table \ref{tab3}), is feasible. The following formula gives the expected luminosity fraction of the third companion:

\begin{equation}
L_{3, O-C}(\%)=\frac{100\cdot
M_{3,min}^{3.5}}{L_1+L_2+M_{3,min}^{3.5}}
\end{equation}

Alternatively, since a cool component of an EB is potentially able to present magnetic activity, one may consider the variation of the quadrupole moment $\Delta$Q and examine its possible influence on the observed period changes as suggested by \citet{b48}. According to the equations of \citet{b49} and \citet{b50}, respectively:
\begin{eqnarray}
\Delta P &=& A \sqrt{2 (1 - \cos \{2\pi P/P_{3}\})},\\
&&\nonumber \\
\frac{\Delta P}{P} &=& -9 \frac{\Delta Q}{Ma^2} \,\,\,\
\end{eqnarray}
and the calculated absolute parameters of the possible magnetically active component (see Table \ref{tab3}), one is able to test this mechanism also as an alternative explanation for the cyclic period variations. According to the criterion suggested by \citet{b50}, the cyclic period changes of an EB could be caused by magnetic influences of a component if its quadrupole moment lies between the range $10^{50}<\Delta Q<10^{51}$.


The eclipsing pair DP Cep is described by a semidetached configuration, where the cooler and more evolved component fills its Roche Lobe. Its primary is located at the upper limit of the MS, indicating that its Hydrogen-burning era has almost come to the end. Although we expected, except the LITE curve, a parabola imposition due to the possible mass transfer, the O -- C analysis did not confirmed it. The long gap of observations between 1952-1975, or the slow mass transfer rate seem to be the best explanations for this result. A third light was detected in the LC analysis ($\sim$5\%), while the minimal mass of the hypothetical MS third companion suggests a light contribution of $\sim$2.5\%. This small difference can be easily overcome, by suggesting an inclination of about 65$^\circ$ for the third body's orbit, which yields a mass of $\sim$0.95 $M_\odot$ capable to produce the observed additional light. Another possible explanation for the cyclic period modulation could be the magnetic activity of the secondary component. Its quadrupole moment was found to have a value of 6.6$\times$10$^{50}~g\cdot~cm^2$, therefore its magnetic influences are potentially candidate formers of the period changes. However, the existence of this mechanism was not detected in the LC analysis (e.g. O'Connell effect), thus other methods of verification (e.g. more LCs, spectroscopy, polarimetry or long term variations of minima timings studies) are needed.

AL Gem is in a detached configuration with the hotter component at the edge of MS and the cooler one more evolved. The LC analysis yielded a third light contribution of $\sim$12.5\%, while the expected one for a MS star with minimal mass $\sim$0.28 $M_\odot$ is  about 0.19\%. This discrepancy might be explained by assuming the non-coplanar orbits of the tertiary component and the EB, similarly to DP Cep. Moreover, the angular separation between the third body and the EB was found to have a relatively large value (98 mas), while the magnitude difference turned out as 2.2 mag. On the other hand, a cool photospheric spot was revealed through the LC analysis, but the application of Applegate's mechanism yielded for the secondary component of the system a value of $\Delta$Q$<10^{50}~g\cdot~cm^2$, therefore is not able to explain the period changes. Spectroscopic and probably also interferometric observations, given the high brightness of the system, will certainly solve the mystery of the existence of the third component.

The system FG Gem was found in semidetached status with the secondary component filling its Roche Lobe and consisting of one MS component and an evolved one. Unlike the previous case, the observed and the expected light contribution from a tertiary component are in perfect agreement (L$_3\sim$4.8\%), hence the system can be considered as a triple, with its components in almost coplanar orbits. However, the O -- C analysis did not reveal any
sign of parabolic behaviour (mass transfer-loss indicator), so very probably the mass transfer has not started yet, or its rate is very slow to be detected in a $\sim$80 $yrs$ time span. The value of quadrupole moment was found for the cool component as 2.5$\times$10$^{50}~g\cdot~cm^2$, and can also cause the observed cyclic changes, but this hypothesis need further observations to verify the existence of magnetic activity.

UU Leo is a detached EB, with the hotter component close to the ZAMS limit indicating its young age and the cooler one beyond the TAMS edge, probably in the subgiant era. Excellent agreement was found between L$_{3,LC}$ and L$_{3,O-C}$ resulted in $\sim$11.3\%. However, a $\Delta$Q value of 6.9$\times$10$^{50}g\cdot~cm^2$ was derived for the secondary component, and, in addition, a cool spot presence was also needed for better fit in the LC analysis. Both mechanisms describe well the O -- C data. As in the case of AL Gem, the relatively high value of L$_3$ is very promising to be detected by spectroscopy which is expected to give definite answer for the mechanism forming the orbital period of the eclipsing pair.

The geometrical status of CF Tau was found to be a detached one. Its components are located beyond the TAMS line, while according to their almost equal masses and positions in the M -- R diagram, they seem to follow similar evolutionary tracks. The observed third light contribution yielded as 4.3\%, while a nearby MS star having a mass value of $\sim$0.89 $M_\odot$, as it was found from the O -- C analysis, is capable to produce this additional luminosity (L$_{3,O-C}$=6.5\%). Thus, since these values are both qualitatively and marginally quantitatively in agreement, the triplicity of the system is established. Both components are potential candidates for magnetic activity, but neither the LC revealed such behaviour nor the value of quadrupole moment ($\Delta$Q$<10^{50}~gr\cdot~cm^2$) is able to explain the cyclic period modulation.

AW Vul can be considered as a classical Algol, since its cooler, more evolved and less massive component fills its Roche Lobe, while the hotter one is a typical MS star. The mass transfer mechanism was not detected in the O -- C analysis, but likely to FG Gem it is possible that its rate is very slow to play a significant role in such a relatively small observational time coverage ($\sim$80 $yrs$). No evidence of magnetic activity was found in the LC analysis, and in addition the Applegate's mechanism is insufficient to describe the cyclic variations of the orbital period, since a value of $\Delta$Q$<10^{50}~g\cdot~cm^2$ was yielded for the cooler component. On the other hand, the observed additional light was found as L$_{3,LC}\sim$3\%, while the expected one from a minimal mass of 0.24 $M_\odot$ was L$_{3, O-C}\sim$0.1\%. This difference can be explained either by assuming the non MS nature of the third body and by inserting a giant age hypothesis or that its inclination is $\sim$25$^\circ$ instead of coplanar orbit, which, according to mass function, yields a value of M$_{3}\sim$0.7 $M_\odot$. Such a mass value satisfies the observed third light in the case of a MS star.

Radial velocity measurements are needed for deriving the absolute parameters of these close binaries more conclusively, while in the cases of AL Gem and UU Leo, where a significant third light was traced, the spectroscopic detection of the third body will verify the current results. New times of minima for all systems will help to achieve a better coverage of the cyclic changes of their orbital periods and will probably reveal new variations which are not so obvious in the current O -- C data sets. Moreover, a prospective interferometric detection of the third body would be also very profitable, with the system AL Gem being the most suitable one for such observations.

\section*{Acknowledgments}

This work has been financially supported by the Special Account for Research Grants No 70/4/9709 of the National \& Kapodistrian University of Athens, Hellas for A.L. and P.N. and also by the Czech Science Foundation grant no. P209/10/0715 for P.Z. In the present work, the minima database: (http://var.astro.cz/ocgate/) has been used. The analyses of DP Cep, FG Gem and CF Tau were based on data from the OMC Archive at LAEFF, preprocessed by ISDC
(https://sdc.laeff.inta.es/omc/), ASAS project (http://www.astrouw.edu.pl/asas/) and SWASP project (http://www.wasp.le.ac.uk/public/lc/), respectively. This research has made use of the SIMBAD database, operated at CDS, Strasbourg, France, and of Astrophysics Data System Bibliographic Services (NASA).

\end{document}